\documentclass{article}
\usepackage{amsmath,graphicx,mlspconf,amssymb}
\usepackage{algorithm,algpseudocode}
\usepackage{booktabs}
\usepackage{hyperref}
\usepackage{cleveref}
\toappear{2024 IEEE International Workshop on Machine Learning for Signal Processing, Sept.\ 22--25, 2024, London, UK}

\title{Episodic fine-tuning prototypical networks for optimization-based few-shot learning: application to audio classification}

\name{Xuanyu Zhuang, Geoffroy Peeters, Gaël Richard\thanks{This work was partly funded by the LISTEN-Lab, a research laboratory dedicated to machine listening from Télécom-Paris (\url{https://listen.telecom-paris.fr/en/}).}}

\address{%
   LTCI, Télécom Paris, IP Paris, France \\
   \{xuanyu.zhuang, geoffroy.peeters, gael.richard\}@telecom-paris.fr
}
\begin{document}
\ninept

\copyrightnotice{\begin{minipage}{\textwidth}
\scriptsize\copyright 2024 IEEE. Personal use of this material is permitted. Permission from IEEE must be obtained for all other uses, in any current or future media, including reprinting/republishing this material for advertising or promotional purposes, creating new collective works, for resale or redistribution to servers or lists, or reuse of any copyrighted component of this work in other works.
\end{minipage}}

\maketitle

\begin{abstract}
The Prototypical Network (ProtoNet) has emerged as a popular choice in Few-shot Learning (FSL) scenarios due to its remarkable performance and straightforward implementation. 
Building upon such success, we first propose a simple (yet novel) method to fine-tune a ProtoNet on the (labeled) support set of the test episode of a $C$-way-$K$-shot test episode (without using the query set which is only used for evaluation). 
We then propose an algorithmic framework that combines ProtoNet with optimization-based FSL algorithms (MAML and Meta-Curvature) to work with such a fine-tuning method. 
Since optimization-based algorithms endow the target learner model with the ability to fast adaption to only a few samples, we utilize ProtoNet as the target model to enhance its fine-tuning performance with the help of a specifically designed episodic fine-tuning strategy. 
The experimental results confirm that our proposed models, MAML-Proto and MC-Proto, combined with our unique fine-tuning method, outperform regular ProtoNet by a large margin in few-shot audio classification tasks on the ESC-50 and Speech Commands v2 datasets. 
We note that although we have only applied our model to the audio domain, it is a general method and can be easily extended to other domains.
\end{abstract}
\section{Introduction}
\label{sec:intro}
Few-shot learning (FSL) has gained increasing attention for its ability to achieve robust generalization to novel classes with limited training data, making it popular for various machine learning tasks. 
While these tasks have been largely centered in the domain of images, there is a growing trend in recent years to extend FSL methods to the audio domain, as collecting or annotating audio data can be very costly in many scenarios \cite{birdclef}. 

FSL algorithms work by constructing a $C$-way-$K$-shot support set during inference to assist in classifying query samples in the query set, in which $C$ represents the number of support classes and $K$ represents the number of samples per support class. 
Each support set and query set together form an episode. 
Vinyals \textit{et al.} propose an episodic training method in \cite{matchingnet}, that is to replace each training batch with also a C-way-K-shot episode and train the model to perform exactly the same behavior as in the test scenario. 
This approach prepares the model for inference tasks in advance, leading to improved performance on the test set. 
FSL algorithms can be broadly divided into two categories.
\textit{Metric-based algorithms}, represented by Protypical Network (ProtoNet)~\cite{prototypical}, are primarily used for classification tasks. 
They predict the label of a query sample by comparing it with support samples in a learned embedding space. 
\textit{Optimization-based algorithms} (also known as meta-learning or learning to learn), represented by Model-Agnostic Meta-learning (MAML)~\cite{maml}  aims at enabling an arbitrary model to quickly fit any new task within several gradient steps on only a few samples by finding an appropriate model initialization. 

However, we observe that for metric-based learners like ProtoNet, the labeled support set provided during inference is only used for comparison with query samples, although there is potential of using it to further fine-tune the model. 
Some prior works have attempted this approach \cite{adapte+embedding} \cite{adaptive}, but the improvements in model performance have been very limited. 
In this paper, we successfully achieve a clear improvement in the performance of ProtoNet on Few-Shot audio classification on ESC-50~\cite{esc50} and Speech Commands v2~\cite{speech} datasets by a simple (yet novel) fine-tuning method.

Our contributions are the following:
\begin{itemize}
    \item We introduce a novel method of fine-tuning a ProtoNet on the labeled support sets during inference.
    \item We propose two new algorithms, \textbf{MAML-Proto} and \textbf{MC-Proto} which embed ProtoNet into the respective optimization-based FSL algorithms (MAML and Meta-Curvature (MC)) with an efficient integration of the proposed fine-tuning method. 
    \item We experimentally show the merits of our few-shot learning approach for audio classification on two diverse public datasets. 
\end{itemize}

The rest of the paper is organized as follows: In \cref{sec:related_work}, we discuss related work including previous attempts to 
apply FSL strategies in audio classification problems. In \cref{sec:background}, we provide some background information regarding the models and algorithms included in our work. We then introduce our proposed fine-tuning method and episodic fine-tuning architectures in \cref{sec:proposal}, and present experiments that demonstrate our proposal in \cref{sec:experiments}.

\section{Related Work}
\label{sec:related_work}
Some prior works have already investigated fine-tuning for ProtoNet or combining metric-based and optimization-based FSL algorithms. Wang \textit{et al.} \cite{hybrid} combines metric-based learners including matching network \cite{matchingnet} and ProtoNet \cite{prototypical} with Meta-SGD \cite{metasgd} by equally dividing support sets into two sub-sets during training. But they did not incorporate any specific design for fine-tuning within their system, resulting in less-than-ideal outcomes. Triantafillou \textit{et al.} \cite{metadataset} propose to transform ProtoNet into an equivalent linear classifier so that it can be easily embedded into MAML \cite{maml}. 
More recently, Gogoi \textit{et al.} \cite{adaptive} introduce an idea of fine-tuning ProtoNet by temporarily attaching a linear classifier layer to the ProtoNet encoder to form a classifier that is fine-tuned on the labeled support set of test episodes. Scott \textit{et al.} \cite{adapte+embedding} also attempted to directly divide the labeled support set provided in the test episodes to two sub-sets for fine-tuning the ProtoNet, except that they did not do this division in a rotational manner nor utilize an optimization-based FSL algorithm to prepare the ProtoNet for fine-tuning.

Yet the works mentioned above, along with the majority of other FSL research, have primarily been conducted in the image domain. 
Here we also list some recent works that have extended the success of FSL from the image domain to the audio domain. 
Meta-audio \cite{metaaudio} marks the first benchmark for 1-shot audio classification on a variety of audio datasets. 
Wang \textit{et al.} introduce a comprehensive paradigm for few-shot sound event detection utilizing negative sampling \cite{sed}, as well as a continual learning framework for few-shot audio classification \cite{continual} which can dynamically adapt to novel classes without retraining. 
Since audio samples are often attached to multiple labels at the same time, Cheng \textit{et al.} \cite{multi} proposed to tackle this by constructing multiple support sets corresponding to each of the attached labels. 

\section{Background}
\label{sec:background}

\subsection{Prototypical Network}

Prototypical Network (ProtoNet)~\cite{prototypical} can be roughly regarded as a nearest neighbor classifier \cite{nn} upon a learned embedding space. In an FSL classification scenario, ProtoNet uses an encoder to project support and query samples into embedding representations. 
The prototype $\mathbf{c}_k$ of class $k$ is defined as the mean of the embeddings of the support sample $S_k$: 
\begin{equation}
    \mathbf{c}_k=\frac{1}{\left|S_k\right|} \sum_{\left(\mathbf{x}_i, y_i\right) \in S_k} f_{\boldsymbol{\theta}}\left(\mathbf{x}_i\right)
\end{equation}
where $\left|S_k\right|$ denotes the number of support samples belonging to class $k$,  $\mathbf{x}_i, y_i$ are the feature vector and corresponding label of the $i^{th}$ sample,  $f_{\boldsymbol{\theta}}$ represents the embedding function of the encoder of learnable parameters $\boldsymbol{\theta}$.

The label of a query $\mathbf{x}$ is then assigned based on a softmax over the Euclidean distance between its embedding $f_{\boldsymbol{\theta}}(\mathbf{x})$ and each prototype $\mathbf{c}_k$:
\begin{equation}
    p_{\boldsymbol{\theta}}(y=k \mid \mathbf{x})=\frac{\exp \left(-d\left(f_{\boldsymbol{\theta}}(\mathbf{x}), \mathbf{c}_k\right)\right)}{\sum_{k^{\prime}} \exp \left(-d\left(f_{\boldsymbol{\theta}}(\mathbf{x}), \mathbf{c}_{k^{\prime}}\right)\right)}
\end{equation}
where $d$ stands for the distance function and $k^{\prime}$ represents the set of all classes.

\subsection{Model-Agnostic Meta-learning} 
Model-Agnostic Meta-learning (MAML) \cite{maml} is a typical optimization-based FSL algorithm. MAML works under the assumption that for an arbitrary model (so-called the learner), there exists a set of initial parameters that could enable the learner model of fast adaption over the distribution of all tasks $p(\mathcal{T})$ with only a few available samples. MAML uses the combination of an inner optimization (line 8-12 in \cref{alg:mcp}) and a meta optimization (line 15 in \cref{alg:mcp}) to learn these appropriate initial parameters. Take FSL classification scenario as an example, the learner $f_{\theta}$ first adapts its initial parameters $\theta$ to the labeled samples in the support set of a sampled episode $\mathcal{T}_i$ and then produces $\theta_i^\prime$:
\begin{equation}
    \theta_i^{\prime}=\theta-\alpha \nabla_\theta \mathcal{L}_{\mathcal{T}_i}\left(f_{\theta}\right)
\end{equation}
where $\alpha$ denotes the (inner) learning rate and $\mathcal{L}$ the loss function. 
The adapted parameters $\theta_i^\prime$ is then evaluated on the corresponding query set of episode $\mathcal{T}_i$ and produce gradients with respect to the initial parameters $\theta$. Such gradients are then aggregated over different sampled episodes to perform an update step on $\theta$ with a meta-learning rate $\beta$:
\begin{equation}
    \theta \leftarrow \theta-\beta \nabla_\theta \sum_{\mathcal{T}_i \sim p(\mathcal{T})} \mathcal{L}_{\mathcal{T}_i}\left(f_{\theta_i^{\prime}}\right)
\end{equation}

\subsection{Meta-Curvature}
Meta-Curvature (MC) \cite{meta_curvature} is a powerful variant of MAML that does not only learn the appropriate initial parameters but also a curvature matrix which further transforms the gradient in the inner optimization stage. 

The meta-curvature function $\mathbf{MC}$ is defined by a composition of three meta-curvature matrices $\mathbf{M}_o \in \mathbb{R}^{C_{\text {out}} \times C_{\text {out}}}$, $\mathbf{M}_i \in \mathbb{R}^{C_{\text {in}} \times C_{\text {in}}}$ and $\mathbf{M}_f \in \mathbb{R}^{d \times d}$, in which $C_{\text{out}}$ , $C_{\text{in}}$ , and $d$ are the number of output channels, the number of input channels, and the filter size respectively. A multi-dimensional gradient $\mathcal{G} \in \mathbb{R}^{C_{\text {out}} \times C_{\text {in}} \times d}$ is transformed as shown in the following equation with all the parameters in meta-curvature matrices being learnable:

\begin{equation}
    \mathbf{M C}(\mathcal{G})=\mathcal{G} \times_3 \mathbf{M}_f \times_2 \mathbf{M}_i \times_1 \mathbf{M}_o 
\end{equation}

where $\times_{n}$ represents the n-mode product between tensors and matrices \cite{meta_curvature}. Compared to MAML, MC brings smoother convergence during training, as well as better generalization performance and faster adaption when fine-tuning with only a few samples during inference.

\section{Proposal}
\label{sec:proposal}
In this section, we propose \textbf{MAML-Proto} and \textbf{MC-Proto} algorithms which further improve the performance of ProtoNet on few-shot audio classification tasks, 
\begin{enumerate}
    \item  by introducing RDFT (\textit{Rotational Division Fine-Tuning}), a simple yet novel method to fine-tune a ProtoNet,
    \item and by training the ProtoNet towards fast adaption within two optimisation-based FSL algorithms (MAML \cite{maml} and MC \cite{meta_curvature}) by a specifically-designed training strategy named episodic fine-tuning, integrating RDFT.
\end{enumerate}


\subsection{Fine-tuning a Prototypical Network}
In FSL scenarios, metric-based algorithms such as ProtoNet are provided with test episodes consisting of a support set and a query set during inference time, in which the support set consists of labeled samples. Using these labeled support samples to further fine-tune the model before the evaluation on the test episode appears as a natural choice. 
However, algorithms like ProtoNet are mostly non-parametric models since they are built upon a nearest neighbor classifier, meaning they can not be fine-tuned through conventional ways. To be more specific, a ProtoNet learns the metric space by comparing labeled support samples with labeled query samples. Since it is not possible to collect labeled query samples during inference time, fine-tuning cannot be conducted with labeled support samples only. 

\begin{figure}[t]
    \centering
    \includegraphics[width=7.5cm]{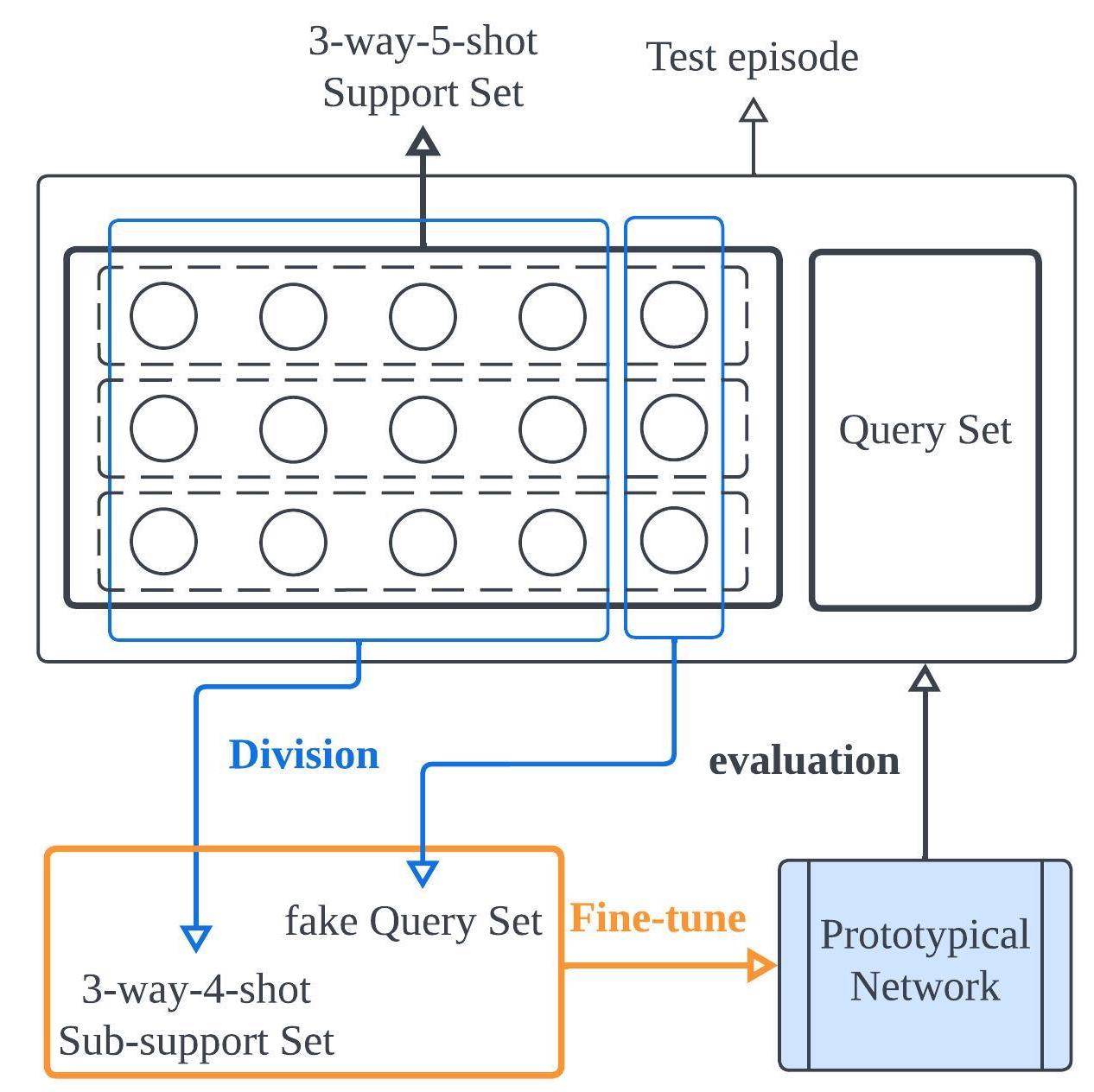}
    \caption{Working mechanism of proposed RDFT method. Here a 3-way-5-shot test episode is used as an example. The 3-way-5-shot labeled support set is divided into a 3-way-4-shot sub-support set and a fake query set, which are together used for fine-tuning the Prototypical Network before evaluating on the original test episode. Here we only present one of the divisions, the blue boxes will move horizontally until every column has been selected as a fake query set.}
    \label{fig:iter_div}
\end{figure}

In this case, we propose a Rotational Division Fine-Tuning method (RDFT), which extracts one labeled sample from each class of the support set on a rotational basis as a fake query set and use the remaining samples as a sub-support set. 
That is, under C-way-K-shot few-shot classification setting, we are therefore creating K different C-way-(K-1)-shot sub-support-set. They can then be used together with the corresponding [C,1] fake query set for fine-tuning the ProtoNet before evaluating on the original C-way-K-shot test episode. The process is illustrated in 
\Cref{fig:iter_div} and a more formal presentation of the method can be found in Algorithm \ref{alg:mcp}.
The motivation behind RDFT is to adhere to the episodic training strategy as closely as possible during the fine-tuning process. For instance, in Figure \ref{fig:iter_div}, the model performs a 3-way-5-shot classification task. While it is not feasible to continue using a 3-way-5-shot task for fine-tuning, RDFT constructs five different 3-way-4-shot sub-tasks to fine-tune the model, thereby approximating the test scenario with maximum effort.

Note that the change of model parameters introduced by RDFT {should not} accumulate across different test episodes during inference time, as it shall only benefit the generalization to novel classes in the current test episode. 
Another point is that RDFT can only work for FSL learning tasks with a number of shots larger than 1.

\subsection{Episodic fine-tuning Prototypical Network within optimiza-tion-based frameworks}

In line with the RDFT method introduced, we further propose to train the ProtoNet within an optimization-based FSL framework. This kind of framework provides the target model with the ability of fast adaption over only a few samples. As for the choice of optimization-based algorithm, we select MC in addition to MAML as it gives the state-of-the-art performance among other optimization-based frameworks on few-shot audio classification tasks \cite{metaaudio}\footnote{For MAML and MC implementations, we used the learn2learn Python package \cite{l2l}.}.
We name them \textbf{MAML-Proto} and \textbf{MC-Proto} respectively, and we emphasize that they work by utilizing MAML / MC to reinforce the adaptive performance of a prototypical network with the help of our proposed novel training strategy called \textit{episodic fine-tuning}.\\

To embed ProtoNet into MAML / MC, an easy way could be to simply replace the training block in the inner optimization stage with a ProtoNet training block under FSL setting. Yet this would conflict with the aforementioned episodic training principle for FSL, 
that is to mimic the test scenario during training. \cite{matchingnet} \cite{lstm}. Since we are fine-tuning the ProtoNet during the test with the RDFT method, we propose to also train the ProtoNet by conducting such fine-tuning in every inner optimization block of MAML / MC. Specifically, the training process can be described by three steps: 
\begin{enumerate}
    \item Randomly sample a $C$-way-$K$-shot episode ${(S_i, Q_i)}$ from the training data, that is to randomly select $C$ classes from the training classes and select $K$ samples from each of the chosen classes. Before evaluating on this episode, the ProtoNet is first fine-tuned from its initial parameters $\theta$ for $n$ steps using RDFT method on only the labeled support set $S_i$.
    \item The fine-tuned ProtoNet with parameters $\theta_i^\prime$ is then evaluated on the same episode ${(S_i, Q_i)}$ and produces a task-specific loss $\mathcal{L}_{(S_i, Q_i)}(f_{\theta_i^{\prime}})$, the so-called \textit{Post-loss}.
    \item Post-loss across different sampled training episodes are aggregated and then together used to update the initial parameters $\theta$ of the ProtoNet with the meta learning rate $\beta$. 
\end{enumerate}

Following these steps, the ProtoNet will optimize towards a better performance after $n$ steps of fine-tuning using RDFT on the support set. 
The detailed algorithmic structure is given in Algorithm \ref{alg:mcp}, taking MAML-Proto as the example.

\begin{algorithm}[t]
\caption{MAML-Proto with Episodic fine-tuning}
\label{alg:mcp}
\begin{algorithmic}[1]
\State \textbf{Require:} Set of sampled episodes $p(\mathcal{E})$
\State \textbf{Require:} Number of fine-tuning steps $n$
\State \textbf{Require:} Inner learning rate $\alpha$, meta learning rate $\beta$
\State Randomly initialize ProtoNet parameters $\theta$
\While{not done}
    \ForAll{${(S_i, Q_i)} \sim p(\mathcal{E})$}
        \State $\theta_i^\prime \leftarrow$ clone $\theta$ \Comment{Start of inner optimization}
        \For{step = $1$ to $n$} \Comment{Start of RDFT}
            \For{j = $0$ to $size(S_i)$} 
                \State $S_i^\prime \leftarrow \{S_{i_{:j}}, S_{i_{j+1:}}\}, \hspace{2mm} Q_i^\prime \leftarrow \{S_{i_j}\}$
                \State Update $\theta_i^{\prime} \leftarrow \theta_i^{\prime}-\alpha \nabla_\theta \mathcal{L}_{(S_i^\prime, Q_i^\prime)}\left(f_{\theta_i^{\prime}}\right)$
                \EndFor{ j}
            \EndFor{ step} \Comment{End of RDFT}
        \EndFor{ ${(S_i, Q_i)}$} \Comment{End of inner optimization}
        \State Update $\theta \leftarrow \theta-\beta \nabla_\theta \sum_{{(S_i, Q_i)} \sim p(\mathcal{E})} \mathcal{L}_{(S_i, Q_i)}\left(f_{\theta_i^{\prime}}\right)$
\EndWhile
\end{algorithmic}
\end{algorithm}

Wang \textit{et al.} adopted a similar idea in \cite{hybrid} of combining metric-based and optimization-based FSL algorithms, but they did not perform any 
fine-tuning. This is indeed suboptimal 
since the optimisation-based FSL frameworks are originally designed for fast and reliable fine-tuning with only few samples available for a novel classes. This could explain the lower performance of their combined algorithms compared to pure metric-based algorithms. On the contrary, our work directly focuses on utilizing the optimization-based FSL algorithms as tools for enhancing the fine-tuning performance of ProtoNet.

\section{Experiments}
\label{sec:experiments}
In this section, we evaluate MAML-Proto and MC-Proto on 5-way-5-shot audio classification tasks using two datasets from different audio domains. 
The goal of these experiments is to demonstrate our model's ability: (1) to achieve fast adaption (i.e. observe an improvement of performance after fine-tuning), 
(2) and to outperform regular ProtoNet.

\subsection{Datasets}

\textbf{ESC-50} is a dataset of environmental sounds including (but not limited to) classes such as animal sounds, urban noise, water sound, etc \cite{esc50}. 
ESC-50 contains 50 classes each represented by 40 samples of 5~sec duration. 
The dataset thus has a large number of classes with few samples per class. 
These features align with FSL settings well, making ESC-50 a popular benchmark for few-shot audio classification task. 

\noindent \textbf{Speech Commands v2} \cite{speech}, on the contrary, is a dataset with fewer classes but a much larger set of samples per class. 
It contains  105,829 1~sec recordings of 35 English command words spoken by different speakers. On a structural level, we infer that Speech Commands v2 is not a very suitable dataset for FSL scenario, as its low class count limits the ability of trained FSL models for novel class generalization. The reason for selecting Speech Commands v2 as a benchmark is to validate whether our proposed methods still work in such a limited situation.

\subsection{Training Setting}
All the experiments are carried under 5-way-5-shot setting. For both datasets, we convert the provided waveform data to Log-Mel-Spectrograms with a sampling rate of 44.1kHz and 16kHz for respectively ESC-50 and Speech Commands v2. A 4-convolutional-block CNN structure is used across all experiments on both of the datasets, with each of the convolutional block consisting of a 64 channel $3\times3$ convolution followed by batch normalization, ReLU, and a $2\times2$ max-pooling layer.

For ESC-50, we set $\alpha$ (the fine-tuning learning rate in the inner-optimization) to 0.2 and the meta learning rate to $1 \times 10^{-3}$. 
For Speech Commands v2, $\alpha$ is set to a lower value, 0.02, and $\beta$ to $1 \times 10^{-5}$. 
The models are fine-tuned for 8 gradient steps on each sampled 5-way-5-shot episode during both training and inference phases for all experiments.

Code examples are available at \url{https://github.com/zdsy/proto-MAML}.

\subsection{Results and Analysis}
\label{sec:4.3}
For ESC-50 and Speech Commands v2 experiments, a regular ProtoNet is used as a baseline to verify whether the proposed models can outperform it. The Average Classification Accuracies (ACC) are calculated on the test set for all models with and without RDFT. This is to investigate the differential impact of employing such a fine-tuning method directly on regular ProtoNet versus on our specifically designed architectures. Additionally, for ESC-50, we selected two additional baselines 
namely ProtoNet+att-sim \cite{attproto} and Proto-HA \cite{hybridatt}. 

\begin{table}[t]
        \centering
        \begin{tabular}{c|c|c}
             \toprule
             &\textbf{w/o} fine-tuning&\textbf{w/} fine-tuning  \\
             \midrule
             ProtoNet&83.95 $\pm$ 0.99\%&66.50 $\pm$ 1.21\%\\
             ProtoNet+att-sim \cite{attproto}&87.7\%&-- \\
             Proto-HA \cite{hybridatt}&\textbf{90.35 $\pm$ 0.83\%}&-- \\
             \textbf{MAML-Proto (Ours)}&83.28 $\pm$ 0.82\%&85.28 $\pm$ 0.87\%\\
             \textbf{MC-Proto (Ours)}&$86.92\pm0.90\%$&\textbf{88.36 $\pm$ 0.87}\%\\
             \bottomrule
        \end{tabular}
        \caption{Average classification accuracy (with 95\% confidence interval) of 5-way-5-shot audio classification task on ESC-50.}
\label{tab:esc501}
\end{table}

\begin{table}[t]
        \centering
        \begin{tabular}{c|c|c}
        \toprule
            &\textbf{w/o} fine-tuning&\textbf{w/} fine-tuning  \\
            \midrule
            ProtoNet&$84.10\pm1.01$\%&$38.66\pm1.37\%$ \\
            \textbf{MAML-Proto (Ours)}&$86.12\pm0.96\%$&$86.28\pm0.92\%$ \\
            \textbf{MC-Proto (Ours)}&\textbf{87.66 $\pm$ 0.91\%}&\textbf{87.74 $\pm$ 0.91}\%\\
            \bottomrule
        \end{tabular}
        \caption{Average classification accuracy (with 95\% confidence interval) of 5-way-5-shot audio classification on Speech Commands v2.}
\label{tab:scv2..}
\end{table}

From the experimental results on ESC-50 and Speech Commands v2 respectively provided in Table \ref{tab:esc501} and Table \ref{tab:scv2..}, we can easily conclude that directly applying RDFT on a regular ProtoNet is actually harmful (see \cref{sec:effect} for explanation). 
On the contrary, conducting such fine-tuning on our proposed frameworks does 
clearly enhance model performance. 
As for MC-Proto, even without fine-tuning, the performance of the model already outperforms that of a regular ProtoNet on both datasets. 
This demonstrates that using episodic fine-tuning training strategy to integrate ProtoNet and MAML / MC already yields enough improvements. 
And on this basis, fine-tuning the models on the support set of a test episode can further optimize the test performance to a very significant extent.

However, the performances on ESC-50 of our best model (MC-Proto) are still lower than those of Proto-HA which is, to the best of our knowledge, the State-Of-The-Art (SOTA) among ProtoNet variants of 5-way-5-shot audio classification tasks on ESC-50. Though, Proto-HA incorporates an additional, rather complex,  hybrid-attention module while our models combining regular ProtoNet with MAML / MC does not introduce any extra parameters.
Nevertheless, we believe that our proposed episodic fine-tuning approach, due to its versatility, can also be readily applied to the Proto-HA algorithm and potentially lead to enhanced performance.

\subsection{The effect of fine-tuning on regular ProtoNet}
\label{sec:effect}

\begin{figure}[t]
\begin{minipage}[b]{1.0\linewidth}
  \centering
  \centerline{\includegraphics[width=8.5cm]{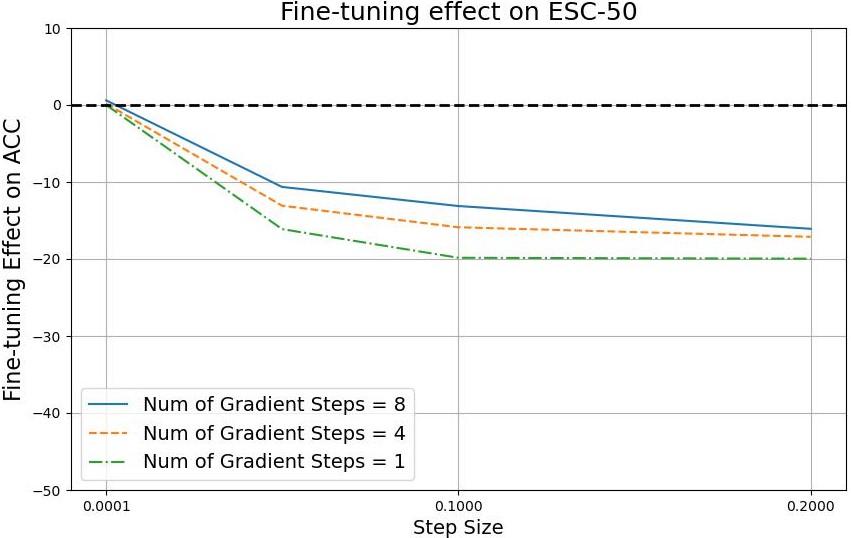}}
  \centerline{(a) Result on ESC-50}\medskip
\end{minipage}
\begin{minipage}[b]{1.0\linewidth}
  \centering
  \centerline{\includegraphics[width=8.5cm]{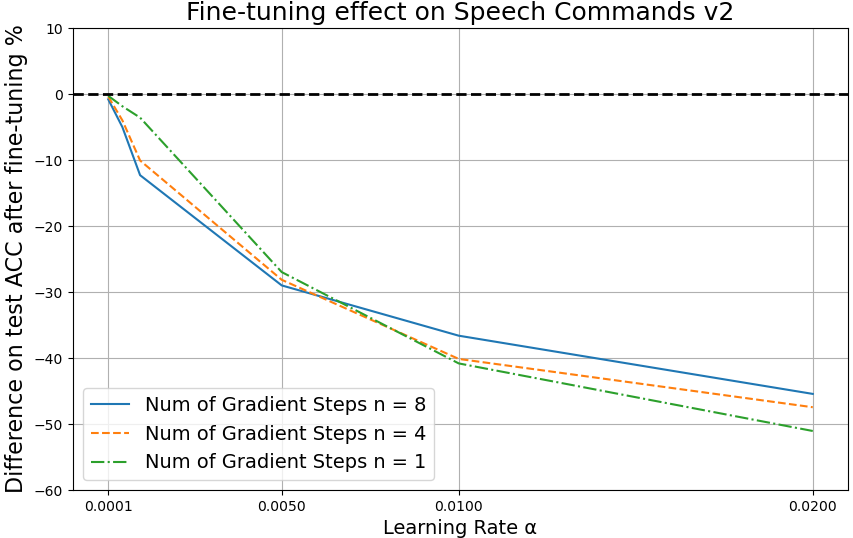}}
  \centerline{(b) Results on Speech Commands v2}\medskip
\end{minipage}
\caption{The change in test ACC of regular ProtoNet after fine-tuning (ACC with fine-tuning - ACC w/o fine-tuning) with different combinations of learning rates and number of gradient steps. The black dotted line represents where the fine-tuning effect on ACC is 0.}
\label{fig:finetune}
\end{figure}
To ensure that the results are more informative, we maintained consistent hyper-parameter settings for fine-tuning experiments on both regular ProtoNet and our proposed models. For instance, on ESC-50, we consistently conduct $8\times5$ gradient steps using RDFT method with a fixed learning rate $\alpha$ of 0.2 for fine-tuning all models. 
Such large fine-tuning amplitude has been proven in \cite{maml} and \cite{metaaudio} to be beneficial for MAML-like algorithms but will probably further increase the risk of overfitting for regular models. 
Therefore, we hypothesize that the severe performance drop of regular ProtoNet caused by RDFT may be attributed to the large fine-tuning amplitude. In other words, the original model parameters with this setting are over adapted neglecting the past learned knowledge.

We then speculate whether reducing the amplitude of RDFT could actually optimize a regular ProtoNet. 
To investigate this, we conducted experiments on both datasets changing the learning rate $\alpha$ and number of gradient steps $n$ to fine-tune regular ProtoNet.
We select several learning rates $\alpha$ ranging from $1\times10^{-4}$ to 0.2 for ESC-50 and $1\times10^{-4}$ to 0.02 for Speech Commands v2, combining with $n=$ 1, 4 and 8 gradient steps to construct the hyper-parameter grid. 
Results are given in \Cref{fig:finetune}. 
From Figure \ref{fig:finetune}.(a) we can observe that for ESC-50 dataset, most of the settings resulted in a performance drop except when step size is set to a very small value. 
Figure \ref{fig:finetune}.(b) demonstrates that RDFT consistently results in a performance decrease for Speech Commands v2 dataset regardless of the hyper-parameter combination used. Furthermore, the performance decrease becomes more pronounced as the amplitude of fine-tuning increases for both datasets.

These observations validate our hypothesis that the performance degradation of a regular ProtoNet brought by RDFT, as shown in \Cref{tab:esc501} and \Cref{tab:scv2..}, is caused by the excessive amplitude of fine-tuning. 
And for ESC-50, reducing the fine-tuning amplitude can indeed make RDFT beneficial for a regular ProtoNet. 
Conversely, the overfitting persists throughout the experiments (even with a very small step size) for Speech Commands v2 and the performance degradation is much more sensitive to changes in step size compared to that on ESC-50. 
We hypothesise that this is likely because the length of audio clips in Speech Commands v2 (1-second) is much shorter than those in ESC-50 (5-second), thus containing less information per sample and resulting in a greater tendency of overfitting. 
This also explains why the performance improvement after fine-tuning for MAML-Proto and MC-Proto on Speech Commands v2 are not as significant as those observed on ESC-50 according to \Cref{tab:esc501} and \Cref{tab:scv2..}. 

Overall, these observations demonstrate that the harmful effects of directly applying RDFT on a regular ProtoNet are attributed to the excessive fine-tuning amplitude, which can be alleviated (or even made beneficial on ESC-50) by reducing the step size and the number of gradient steps used for fine-tuning. Additionally, this indirectly validates the enhancement of large-amplitude fine-tuning performance brought by our proposed episodic fine-tuning models.

\section{Conclusion and future work}
We first introduce a simple yet novel fine-tuning method for ProtoNet named RDFT, that is to divide the labeled support set provided during the test into a sub-support set and a fake query set in a rotational manner, which are then used to fine-tune the ProtoNet before evaluating on the original test episode. To cooperate with such a fine-tuning method and enhance the ProtoNet's ability of fast adaption, we propose to further embed ProtoNet into MAML and MC with the use of episodic fine-tuning strategy. 
The experiments conducted on ESC-50 and Speech Commands v2 datasets demonstrate that combining RDFT with MAML-Proto and MC-Proto algorithms significantly surpasses the performance of a regular ProtoNet in few-shot audio classification tasks. 
This improvement persists even though such a large-amplitude fine-tuning using RDFT has been proven harmful when used alone on regular ProtoNet, further highlighting our models' ability of fast and efficient adaption.

As for future work, one promising direction is to extend our algorithm to other metric-based FSL algorithms including Matching Network \cite{matchingnet}, Relation Network \cite{relation} and attention-based models like Proto-HA \cite{hybridatt}. Others include to further implement our models on tasks with different number of ways and shots, and to provide comprehensive theoretical support for RDFT method. Future experiments will also be conducted on a larger variety of datasets in different domains to further investigate the merits and robustness of our approach.




\bibliographystyle{IEEEbib}
\bibliography{refs}

\begin{thebibliography}{10}

\bibitem{birdclef}
Stefan Kahl, Amanda Navine, Tom Denton, Holger Klinck, Patrick Hart, Herv{\'e} Glotin, Herv{\'e} Go{\"e}au, Willem-Pier Vellinga, Robert Planqu{\'e}, and Alexis Joly,
\newblock ``Overview of birdclef 2022: Endangered bird species recognition in soundscape recordings.,''
\newblock in {\em CLEF (Working Notes)}, 2022, pp. 1929--1939.

\bibitem{matchingnet}
Oriol Vinyals, Charles Blundell, Timothy Lillicrap, Daan Wierstra, et~al.,
\newblock ``Matching networks for one shot learning,''
\newblock {\em Advances in neural information processing systems}, vol. 29, 2016.

\bibitem{prototypical}
Jake Snell, Kevin Swersky, and Richard Zemel,
\newblock ``Prototypical networks for few-shot learning,''
\newblock {\em Advances in neural information processing systems}, vol. 30, 2017.

\bibitem{maml}
Chelsea Finn, Pieter Abbeel, and Sergey Levine,
\newblock ``Model-agnostic meta-learning for fast adaptation of deep networks,''
\newblock in {\em International conference on machine learning}. PMLR, 2017, pp. 1126--1135.

\bibitem{adapte+embedding}
Tyler Scott, Karl Ridgeway, and Michael~C Mozer,
\newblock ``Adapted deep embeddings: A synthesis of methods for k-shot inductive transfer learning,''
\newblock {\em Advances in Neural Information Processing Systems}, vol. 31, 2018.

\bibitem{adaptive}
Manas Gogoi, Sambhavi Tiwari, and Shekhar Verma,
\newblock ``Adaptive prototypical networks,''
\newblock {\em arXiv preprint arXiv:2211.12479}, 2022.

\bibitem{esc50}
Karol~J Piczak,
\newblock ``Esc: Dataset for environmental sound classification,''
\newblock in {\em Proceedings of the 23rd ACM international conference on Multimedia}, 2015, pp. 1015--1018.

\bibitem{speech}
Pete Warden,
\newblock ``Speech commands: A dataset for limited-vocabulary speech recognition,''
\newblock {\em arXiv preprint arXiv:1804.03209}, 2018.

\bibitem{hybrid}
Duo Wang, Yu~Cheng, Mo~Yu, Xiaoxiao Guo, and Tao Zhang,
\newblock ``A hybrid approach with optimization and metric-based meta-learner for few-shot learning,'' 2019.

\bibitem{metasgd}
Zhenguo Li, Fengwei Zhou, Fei Chen, and Hang Li,
\newblock ``Meta-sgd: Learning to learn quickly for few-shot learning,''
\newblock {\em arXiv preprint arXiv:1707.09835}, 2017.

\bibitem{metadataset}
Eleni Triantafillou, Tyler Zhu, Vincent Dumoulin, Pascal Lamblin, Utku Evci, Kelvin Xu, Ross Goroshin, Carles Gelada, Kevin Swersky, Pierre-Antoine Manzagol, et~al.,
\newblock ``Meta-dataset: A dataset of datasets for learning to learn from few examples,''
\newblock {\em arXiv preprint arXiv:1903.03096}, 2019.

\bibitem{metaaudio}
Calum Heggan, Sam Budgett, Timothy Hospedales, and Mehrdad Yaghoobi,
\newblock ``Metaaudio: A few-shot audio classification benchmark,''
\newblock in {\em International Conference on Artificial Neural Networks}. Springer, 2022, pp. 219--230.

\bibitem{sed}
Yu~Wang, Justin Salamon, Nicholas~J Bryan, and Juan~Pablo Bello,
\newblock ``Few-shot sound event detection,''
\newblock in {\em IEEE International Conference on Acoustics, Speech and Signal Processing (ICASSP)}, 2020, pp. 81--85.

\bibitem{continual}
Yu~Wang, Nicholas~J Bryan, Mark Cartwright, Juan~Pablo Bello, and Justin Salamon,
\newblock ``Few-shot continual learning for audio classification,''
\newblock in {\em IEEE International Conference on Acoustics, Speech and Signal Processing (ICASSP)}, 2021, pp. 321--325.

\bibitem{multi}
Kai-Hsiang Cheng, Szu-Yu Chou, and Yi-Hsuan Yang,
\newblock ``Multi-label few-shot learning for sound event recognition,''
\newblock in {\em IEEE 21st International Workshop on Multimedia Signal Processing (MMSP)}, 2019, pp. 1--5.

\bibitem{nn}
Thomas Cover and Peter Hart,
\newblock ``Nearest neighbor pattern classification,''
\newblock {\em IEEE transactions on information theory}, vol. 13, no. 1, pp. 21--27, 1967.

\bibitem{meta_curvature}
Eunbyung Park and Junier~B Oliva,
\newblock ``Meta-curvature,''
\newblock {\em Advances in neural information processing systems}, vol. 32, 2019.

\bibitem{l2l}
S{\'e}bastien~MR Arnold, Praateek Mahajan, Debajyoti Datta, Ian Bunner, and Konstantinos~Saitas Zarkias,
\newblock ``learn2learn: A library for meta-learning research,''
\newblock {\em arXiv preprint arXiv:2008.12284}, 2020.

\bibitem{lstm}
Sachin Ravi and Hugo Larochelle,
\newblock ``Optimization as a model for few-shot learning,''
\newblock in {\em International conference on learning representations}, 2016.

\bibitem{attproto}
Szu-Yu Chou, Kai-Hsiang Cheng, Jyh-Shing~Roger Jang, and Yi-Hsuan Yang,
\newblock ``Learning to match transient sound events using attentional similarity for few-shot sound recognition,'' 2019.

\bibitem{hybridatt}
You Wang and David~V. Anderson,
\newblock ``Hybrid attention-based prototypical networks for few-shot sound classification,''
\newblock in {\em IEEE International Conference on Acoustics, Speech and Signal Processing (ICASSP)}, 2022, pp. 651--655.

\bibitem{relation}
Flood Sung, Yongxin Yang, Li~Zhang, Tao Xiang, Philip~HS Torr, and Timothy~M Hospedales,
\newblock ``Learning to compare: Relation network for few-shot learning,''
\newblock in {\em Proceedings of the IEEE conference on computer vision and pattern recognition}, 2018, pp. 1199--1208.

\end{thebibliography}

\end{document}